# Ultra-stable optical frequency dissemination on a multi-access fibre network

Anthony Bercy,[1,2] Olivier Lopez,[1] Paul-Eric Pottie[2] and Anne Amy-Klein[1,*]

[1]Laboratoire de Physique des Lasers, Université Paris 13, Sorbonne Paris Cité, CNRS, 99 Avenue Jean-Baptiste Clément, 93430 Villetaneuse, France
[2]Laboratoire National de Métrologie et d'Essais–Système de Références Temps-Espace, UMR 8630 Observatoire de Paris, CNRS, UPMC, 61 Avenue de l'Observatoire, 75014 Paris, France
*Corresponding author: amy@univ-paris13.fr

**Abstract**
We report the dissemination of an ultrastable optical frequency signal to two distant users simultaneously using a branching network. The ultrastable signal is extracted along a main fibre link; it is optically tracked with a narrow-linewidth laser diode, which light is injected in a secondary link. The propagation noise of both links is actively compensated. We implement this scheme with two links of 50-km fibre spools, the extraction being setup at the mid-point of the main link. We show that the extracted signal at the end of the secondary link exhibits fractional frequency instability of $1.4 \times 10^{-15}$ at 1-s measurement time, almost equal to the $1.3 \times 10^{-15}$ instability of the main link output end. The long-term instabilities are also very similar, at a level of $3\text{-}5 \times 10^{-20}$ at $3 \times 10^4$-s integration time. We also show that the setting up of this extraction device, or a simpler one, at the main link input, can test the proper functioning of the noise rejection on this main link. This work is a significant step towards a robust and flexible ultra-stable network for multi-users dissemination.

1. Introduction

Optical fibre links enables us to transfer an ultrastable frequency between distant laboratories with almost no degradation [1, 2]. Since ten years, they have been extended from a few hundreds of km to a record length of 1840-km [3] and noise compensation has been improved leading to an accuracy of the frequency transfer in the $10^{-20}$ range [4]. It is the best mean for comparing distant optical clocks as recently demonstrated on a 1415-km fibre distance between the French and German National Metrological Institutes in Paris and Braunschweig [5]. Behind fundamental frequency metrology, optical links open new applications such as high-precision measurements of atomic or molecular absolute frequencies [6-9], or frequency dissemination to radio-antenna in radioastromy [10].

However, the development of optical links is now slowed down by the difficulty and cost to access the fibres. Even if optical links were also implemented on active telecommunication fibres [11, 12], by using wavelength division multiplexing, a smart management of the fibre network is necessary to extend the dissemination to many laboratories. In that perspective, a point-to-point distribution requires too many fibres and a branching network is much more effective. It was first proposed and demonstrated to extract the ultrastable signal along a main link to multiple users [13-16]. Another possibility consists in implementing a branching optical fibre network with noise correction at each output end [17]. Here, following first proposal for radio-frequency and optical frequency multi-user dissemination [13, 15], we demonstrate the implementation of a secondary link, injected with a laser diode phase-locked to the signal extracted on a main optical link. This new architecture enables us to disseminate an ultrastable signal from a single laboratory to many distant users, who are not necessary along the main link. It is especially suitable for multi-users dissemination in a metropolitan area network.

In this paper, we will first explain the principle of multi-user dissemination using extraction and a secondary link. We will then describe the experimental set-up which uses a main and a secondary link of 50-km fibre spools. We will show the experimental results for the instability and the accuracy of the extracted signal, which is copying the input signal with a very high fidelity. Finally we will show that such a device can be used to test the proper functioning of the main link, as an alternative to the usual end-to-end measurement.

2. Principle

Fig. 1 shows the principle of the multiple dissemination to distant users from a main link, which was first proposed in [15] for radio-frequencies and in [13, 14] for optical frequencies. For clarity, we will first consider the distribution of an ultrastable signal to two users Out0 and Out1. A noise-compensated optical link is established to user Out0, using the well-known technique of the round-trip noise compensation [1, 2]. Corrections are implemented using an acousto-optic modulator (AOM) at the link input. Instead of establishing a second optical link to user Out1, part of the main link is used to disseminate the signal to an intermediate point, called extraction point, where a coupler is inserted to extract both the forward and backward signals propagating in the main link, of frequency $\nu_+$ and $\nu_-$ respectively [13-16]. A narrow bandwidth laser diode is offset phase-locked to the forward extracted signal and feeds a secondary optical link to Out1 [13]. The secondary link is also noise-compensated, using the same method as for the main link. Other secondary links can be fed by the laser diode which enables us to distribute the ultrastable signal to many users simultaneously. A second laser diode can also be phase-locked to the backward extracted signal and feed other secondary links. In that case, the noise compensation will use acousto-optics modulator of opposite shift compared to the ones used for the links fed by the first laser diode [18].

That scheme includes an extraction stage along the main link, as initially proposed in [16] and demonstrated in [13-15, 18]. Both forward and backward extracted signals exhibits phase fluctuations, because the noise correction at the main link input is overcorrecting the phase noise at this intermediate point. Let consider that the extraction is performed at a distance $L_A$ from the input end, with the integrated phase noise $\phi_A$, and $L_B$ from the output end, with the integrated phase noise $\phi_B$ (see Fig. 1). The phase fluctuations on the forward extracted signal are given by the sum of the noise $\phi_A$ and the main link correction signal $\phi_{C0} = -(\phi_A + \phi_B)$, which results in a noise $-\phi_B$. To compensate it, we detect the beat-note of the two extracted signals, which exhibits phase fluctuations $2\phi_B$. The latter signal, after division by two, is mixed with the control signal of the laser diode, which phase fluctuations are thus free of the main link phase noise.

This set-up enables us to distribute the ultrastable signal to many remote users simultaneously along a main link. It is compatible with optical amplification on main or secondary link and thus with distribution to sites more than 100-km away. We will see in next paragraph that its implementation does not require that the local RF oscillator is stable and accurate and it can be fully automatized. It is very flexible, since one additional extraction coupler and/or one additional secondary link can be set up on the main link without changing the overall dissemination architecture. By contrast with the branching architecture proposed in [17], the main link is fully independent of the secondary links. But the noise rejection at the secondary links output depends on the proper functioning of the main link.

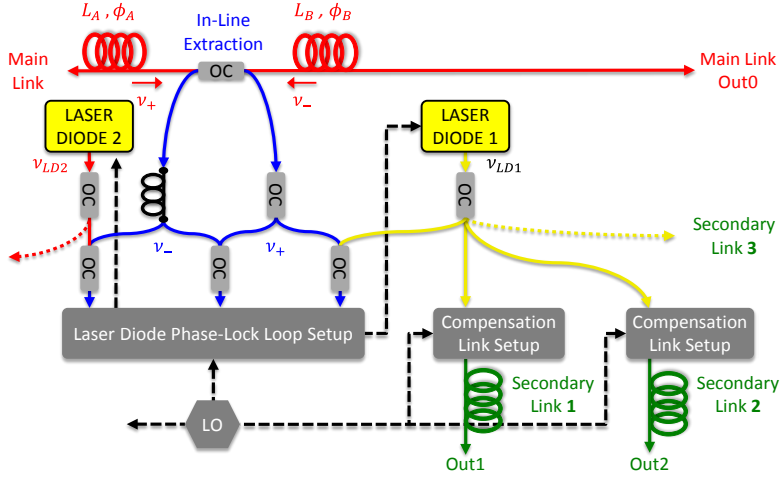

Fig. 1: principle of the multi-user distribution; a laser diode is phase-locked onto the ultrastable signal extracted along the main link and feeds secondary links, which noise are compensated; OC: optical coupler, LO: local oscillator.

3. Experimental set–up description.

The secondary link set-up was already introduced in [13] and is shown in Fig. 2. It is composed of three parts, the extraction set-up, the optical regeneration with the laser diode, and the noise compensation of the secondary link, respectively labelled 1 to 3 in Fig. 2.

The main link is set-up with the usual round-trip noise compensation method [1, 11], corrections being applied at the link input end through an AOM of frequency $f_1$. It is fed with an ultrastable laser of frequency $\nu_0$ which is disseminated from SYRTE through a free-running optical link of 43-km, leading to frequency instabilities in the range of $10^{-14}$ [19].

Part of the forward and backward signals propagating on the main link are extracted with a coupler and recombined with a second coupler (part 1 in blue in Fig. 2). This Mach-Zehnder type interferometer gives a beatnote signal on a first photodiode PD1 at frequency $2 \times f_2$, where $f_2$ is the frequency of the AOM at the main link output. This signal exhibits the noise $2\phi_B$ of the upstream part of the main link. A polarisation controller is used to align the polarisation of the two beams, which are perpendicular at the extraction coupler output, since a Faraday mirror is used to reflect the signal at the end of the main link. A 90° Faraday rotator could also be used.

The beatnote signal between the laser diode and the forward extracted signal is detected with a second photodiode PD2 (part 2 in yellow in Fig. 2). A polarisation controller at the laser diode output (not shown in Fig. 2) enables us to optimize its amplitude. This beatnote signal exhibits both the laser diode phase fluctuations $\phi_{LD}$ and the main link propagation noise $-\phi_B$. The latter noise is rejected by mixing this beatnote signal with half the beatnote signal detected on PD1. The resulting signal is used to phase lock the laser diode with an offset frequency $f_{LD}$ given by the local oscillator frequency $f_{LO}$. The corrections are applied to both the laser temperature and current, with a bandwidth about 100 kHz. The laser diode frequency is thus $\nu_{LD} = \nu_0 + f_1 + f_2 + f_{LO}$ and its phase fluctuations are copying the phase fluctuation $\phi_{LO}$ of the local oscillator (within the loop bandwidth).

The laser diode is feeding the secondary optical link which noise $\phi_S$ is compensated with again the same round-trip method (part 3 in green in Fig. 2). Two AOMs (denoted AOM3 and AOM4) of negative frequency shifts $f_3$ and $f_4$ are used at the input and output of this secondary link, respectively. The beatnote between the round-trip signal and the input signal, of frequency $2\times(f_3 + f_4)$, is detected through a Michelson-type interferometer. It exhibits the phase noise $-2\phi_S$ (the minus sign arises from the negative frequency shifts). After division by 30, it is mixed with the signal of the local oscillator, itself divided by 15, to generate corrections applied through AOM3, with a frequency $f_3 = f_{LO} - f_4$ and a phase $\phi_{C1} = -(\phi_S + \phi_{LO})$. Thus the secondary link output frequency is copying the ultra-stable frequency at the main link output $\nu_0 + f_1 + f_2$ without any added phase noise.

Although using a local RF oscillator for the phase-lock loops and AOM drivers, this set-up is insensitive to its frequency fluctuations. We have indeed engineered the set-up so that the RF oscillator instabilities and bias added with the laser diode phase-lock loop are exactly compensated by the secondary link phase corrections. This passive noise compensation is similar to that implemented in the laser stations we have developed for cascaded optical links [12].

Furthermore we designed this set-up to minimize the noise resulting from uncompensated fibre paths [13]. The fibre lengths are minimised and all the passive optical set-up (including the couplers and interferometers) is very compact and installed in an aluminium box which is actively temperature-stabilised at 30° and surrounded with insulating foam. Moreover we carefully adjusted the fibre lengths in order that the temperature sensitivity of the set-up is further reduced. We chose the fibre lengths of both arms between the couplers OC1 and OC2 to be identical (see Fig. 2). In the case of a homogeneous temperature, the phase variation due to any temperature change will thus cancel between the two arms. We also set the fibre length between the laser diode and the optical coupler OC4 to be the sum of the fibre lengths between OC1 and OC4 and between the laser diode output and the Faraday mirror FM1 after OC6. That way, we add to the laser diode phase the opposite of the uncompensated fibre phase variation between the laser output and the Faraday mirror FM1 and the resulting total phase variation cancels.

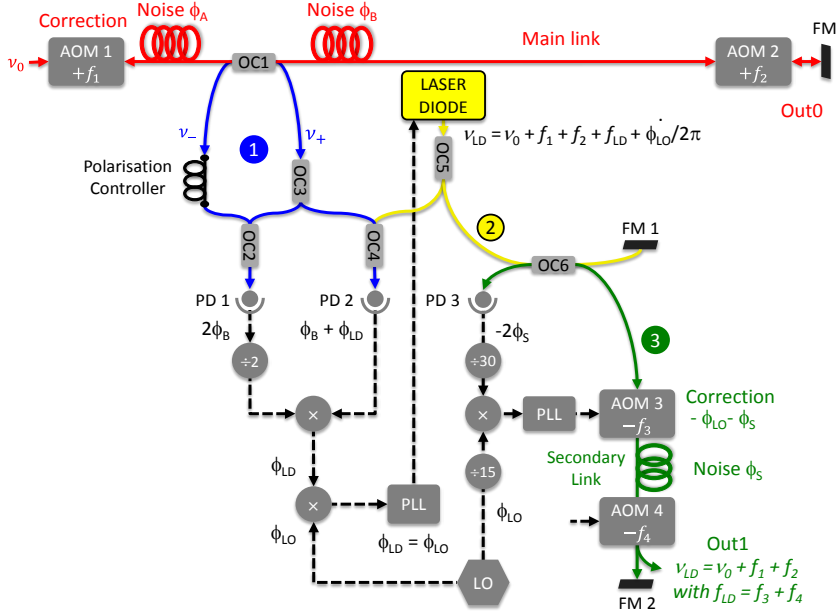

Fig. 2: Set-up for a secondary link with noise compensation; OC: optical coupler, AOM: acousto-optic modulator, PLL: phase-lock loop, FM: Faraday mirror, PD: photodiode, LO: local oscillator. For simplicity, only the phases have been indicated for the PLLs.

4. Results and discussion.

We implement the multi-user distribution scheme of Fig. 2 at the mid-point of an optical link of 50-km fibre spools. The secondary link is 50-km long. We use an interferometric set-up to detect simultaneously the beatnotes between the input end of the link and the output ends of the main and secondary links, respectively labelled Out0 and Out1 in Fig. 2. These beatnotes enable us to characterize the performance of the frequency transfer at the main and secondary links output ends, the latter being denoted by extraction output below.

The phase noise Power Spectral Densities (PSD) of these two signals are plot in Fig. 3 (red curve (c) and green curve (d) respectively). The phase noise PSD of the free-running main link (orange curve (a)) and extraction output (black curve (b)) are also shown. By contrast to the other curves, they were measured with a frequency voltage converter which limits the measurement sensitivity after 1 kHz. Both curves are very similar, and typical of optical fiber links, with a noise around 10 rad$^2$/Hz at 1 Hz and 10$^{-6}$ rad$^2$/Hz at 1 kHz. Both compensated noise are also very similar with a phase noise PSD below 10$^{-5}$ rad$^2$/Hz between 1 and 50 Hz. The noise is corrected up to around 150 Hz, with a correction overshoot of a few hundreds of Hz compatible with the theoretical bandwidth of 1 kHz given by 1/4τ with τ the propagation delay [2]. Note that this limit is the same for both links but, in case the lengths are different, the bandwidth of the extraction is limited by the longer delay [13]. We checked that the noise floors of both outputs are below these PSDs. The noise rejection of around 10$^6$ at 1 Hz is also compatible with the theoretical limit given by $\frac{1}{3}(2\pi f \tau)^2$ which is 8×10$^7$ for a 50-km link [2]. This shows that the noise rejection is optimized.

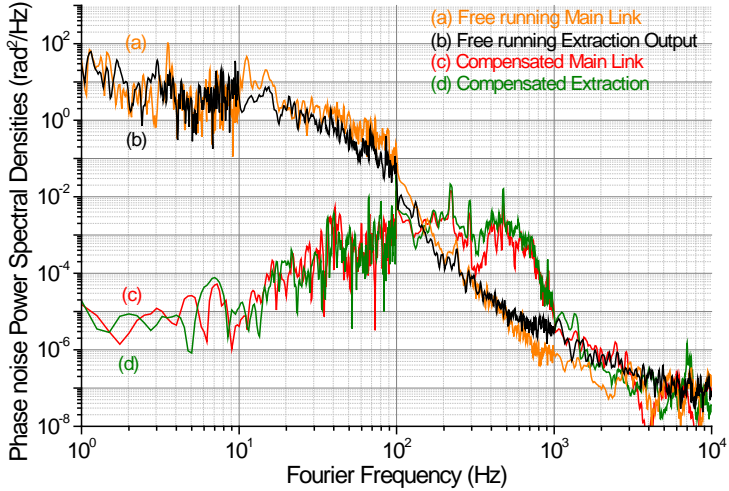

Fig. 3: Phase noise power spectral density of (a) the free-running main link (orange line), (b) the free-running extraction output (black line), (c) the compensated main link (red line) and (d) the compensated extraction output (green line).

To evaluate the end-to-end stabilities, the beat-notes at frequency $f_1 + f_2 = 75$ MHz are tracked with a 100 kHz bandwidth and after frequency division by 2 they are recorded simultaneously with a dead-time free counter (Kramer+Klische FXE) with a gate time of 1 s and Λ-type operation, in order to average out the noise in a narrow bandwidth [20, 21]. We then use the modified Allan deviation to characterise the stability of the extraction set-up.

Fig. 4 displays the experimental stability of the free-running main link (orange circles (a)) and secondary link (black down-triangles (b)). Their noise slightly differs because the main link fibre is wound onto a spool and thus experiences some constraints, whereas the secondary link fibre is wound freely (without any central spool). The extraction output stability (green squares (c)) is $1\times10^{-17}$ at 1 s averaging time, decreases and reaches a floor of about $4\times10^{-20}$ after $10^4$ s. It is at the state of the art for fibre frequency transfer on such distances and nearly copies the main link output stability. Note that the noise correction is very robust and that the set-up was operated during 2.5 day without any cycle-slip. At short-term, the stability is slightly above the main link output stability (red up-triangles (d)) which is $6\times10^{-18}$ at 1 s averaging time. The extraction output indeed exhibits both the contribution of the signal extracted from the main link (after compensation of $\phi_B$) and the residual noise of the secondary link, the latter limiting the noise rejection at short term. In Fig. 4 is also shown the noise floors for both the main link (pink diamonds (f)) and the extraction output (grey stars (e)), which limit the long-term stability. The latter is mainly attributed to thermal noise on uncompensated fibre paths, due to imperfect length adjustment and thermal stabilisation in the extraction optical set-up, or in the interferometric measurement set-up.

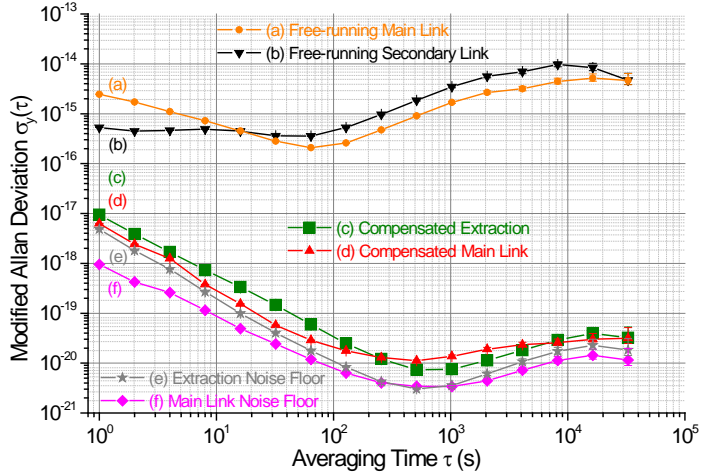

Fig. 4: End-to-end fractional frequency instability, calculated from Λ-type data with the Modified Allan deviation, of (a) the 50-km free-running main link (orange circles), (b) the 50-km secondary link (black down-triangles) (c) the compensated extraction output (green squares), (d) the compensated main link output (red up-triangles), (e) the extraction noise floor (grey stars) and (f) the main link noise floor (pink diamonds).

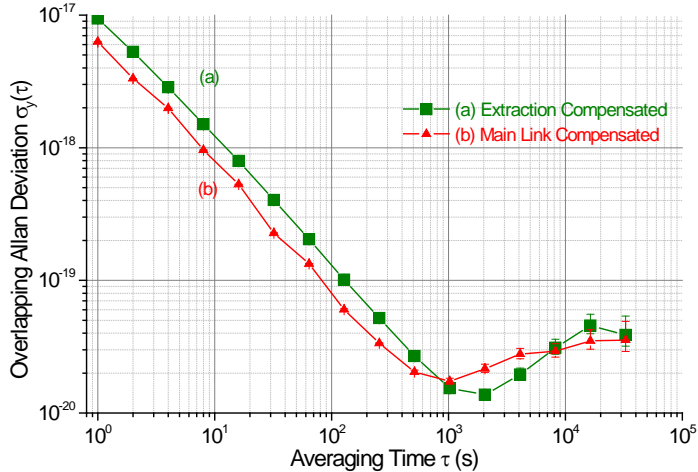

Fig. 5: End-to-end fractional frequency instability, calculated from Λ-type data with the overlapping Allan deviation, of (a) the compensated extraction output (green squares) and (b) the compensated main link output (red up-triangles).

Finally, the accuracy of the frequency transfer was evaluated by calculating the mean value of the end-to-end beat note frequency offset. Following [4, 22], we estimate its statistical fractional uncertainty as the long-term overlapping Allan deviation at 20000 or 30000 s of the data set. We obtained for the main link a mean frequency offset of $-8.2 \times 10^{-21}$ with a statistical uncertainty of $4 \times 10^{-20}$. For the extraction output, the mean frequency offset is $-2.4 \times 10^{-21}$ with a statistical uncertainty of $4 \times 10^{-20}$. These figures show that the frequency transfer shows no bias to the expected value.

With this link performance, the ultra-stable laser is thus transferred through the secondary and main links without degradation. Even a record ultrastable laser exhibiting a $10^{-16}$ stability at 1 s averaging time [23] would be faithfully transferred for integration time above 1 s.

5. Testing the noise rejection on the main link

We have shown that our extraction set-up enables us to transfer the ultrastable signal to several users simultaneously, using secondary links to reach the more distant ones. We are now discussing an alternative application of an extraction set-up, which consists in testing the proper functioning of the noise rejection on the main link.

As pointed out by several authors, one can test the performance of an optical link by analysing the end-to-end stability, which requires that the two ends are at the same place. Thus optical links are usually implemented with two parallel fibres or with a loop fibre. But it is not always possible to get such configuration, which requires the doubling of the fibre lengths to the distant lab, and thus increases the equipment and maintenance costs. Calonico and co-workers implemented a double round-trip on a single fibre in order to characterize the link performance [24]. This alternative method is very beneficial for long-haul link network development but could be difficult to implement when the fibre attenuation and parasitic noise are large. Here we propose to use an extraction set-up at the link input to test the good functioning of the noise rejection on the main link. As demonstrated in [13], the extraction end phase noise PSD without any secondary link is lower or equal to the main link output phase noise PSD. When the extraction occurs at the link input, the phase correction at the extraction output is exactly opposite to the roundtrip fibre phase noise and thus to the correction at the link input. Therefore, following our simple model detailed in [13], the extracted signal copies exactly the input signal and no fibre phase fluctuation is added to the signal. Thus, by analysing the extracted signal, we can infer that the noise correction signal applied at the link input is properly compensating the main link noise. For that purpose, one has to carefully implement the stability measurement set-up: the lengths of the fibres used to detect the beatnote between the input end of the link and the extraction end should be carefully balanced and thermalized in order that the noise arising from these uncommon fibre paths is negligible.

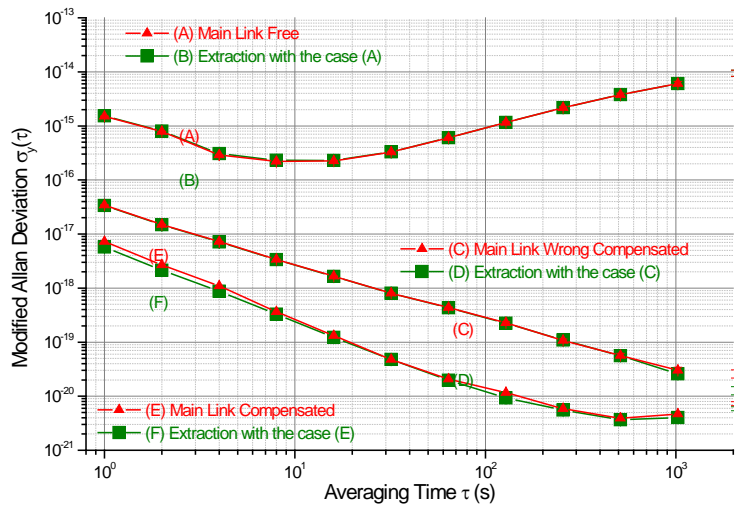

Fig. 6: End-to-end fractional frequency instability, calculated from Λ-type data with the Modified Allan deviation, of the main link (red up-triangles) and the extraction output (green squares) for three pairs of data: (a) and (b) free-running main link, (c) and (d) poorly compensated main link, and (e) and (f) compensated main link. Error bars are too small to show on Figure.

In order to validate this method, we implemented our extraction set-up at the main link input, and we replaced the secondary link by an equivalent attenuator. Then we compare the main link output stability with the extraction stability for the following three cases: the free-running main link, a partial noise compensation of the main link, with a poor adjustment of the loop gains, and an optimized noise compensation of the main link. The three pairs of stabilities are shown in Fig. 6, with red-up triangles for the main link and green squares for the extraction output. In each case, the extraction stability is almost copying the main link output stability. Note that for the compensated main link, the extraction stability is $5.9 \times 10^{-18}$ at 1-s averaging time when the main link output stability is $7.5 \times 10^{-18}$. The ratio of the square of these stabilities gives an estimate of the F-factor we introduced in [13]. It is 0.6, when the theoretically expected value is 0. This can be explained by the fact that the stability at 1 s is limited by the noise floor (as can be seen on Fig. 4) which is expected to be mainly due to the electronic detection set-up.

The stabilities displayed in Fig. 6 show that the extraction stability is a very good measurement of the main link stability and enables us to check the right functioning of the noise rejection. Such set-up can be used to test the proper operation of a single-fibre link. It is very beneficial when implementing the first spans of a cascaded link without having set-up the next spans. It is also very useful when implementing an uncompensated link to a distant user, in order to evaluate these user link instabilities. Note that, for these applications, we don't need to phase-lock a laser diode to the extracted signal; the simpler extraction set-up we have demonstrated in [13] is more than enough.

6. Conclusion

We have demonstrated the distribution of an ultrastable frequency to two distant users using a single link input. We use a new type of branching network topology with a secondary link departing from any position along a main link. This architecture enables us to establish an ultrastable frequency distribution to multiple users, for example in a dense metropolitan area. It is a very interesting alternative to point-to-point distribution, which requires one optical link per user and thus a higher cost of fibre equipment and maintenance. We have shown that the stability at the secondary link output is nearly copying that at the main link output, with an Allan deviation of $10^{-17}$ at 1-s averaging time. This set-up is very robust and can be operated without any stable RF oscillator at the extraction point, since the secondary link output stability is independent of this local oscillator's instabilities. It can be implemented on an active telecommunication network with simultaneous data traffic, provided that we set-up optical add-drop multiplexers to insert or extract the ultrastable signal from the data traffic [11].

This work is a major step of the French project named REFIMEVE+ [25]. REFIMEVE+ is developing a wide national infrastructure where a reference optical signal generated at SYRTE will be distributed to about twenty academic and institutional users using the National Academic network RENATER. Such an extraction set-up will be mainly implemented for metropolitan area distribution, as for example in the Paris area. It is also a key-element of the current effort to establish continental ultra-stable fibre network. Multi-users dissemination opens the way to a wide distribution of an ultra-stable frequency reference, enabling applications beyond metrology [6-10] and we expect that it will stimulate the development of new high-sensitivity experiments in a broad range of applications.

Acknowledgments

We acknowledge Christian Chardonnet and Saïda Guellati-Khelifa for stimulating this work and Giorgio Santarelli for fruitful discussions. We also acknowledge A. Kaladjian and F. Wiotte from LPL for technical support. We acknowledge funding support from the Agence Nationale de la Recherche the Agence Nationale de la Recherche (ANR blanc LIOM 2011-BS04-009-01, Labex First-TF ANR 10 LABX 48 01) and IFRAF-Conseil Régional Ile-de-France.